\documentclass[conference,a4paper]{IEEEtran}
\IEEEoverridecommandlockouts

\usepackage{cite}
\usepackage{amsmath,amssymb,amsfonts,bm}
\usepackage{algorithmic}
\usepackage{graphicx}
\usepackage{textcomp}
\usepackage{xcolor}
\usepackage{commath}
\usepackage{mathbbol}
\usepackage{dblfloatfix}
\usepackage{flushend}
\usepackage{float}
\usepackage{balance}
\usepackage{subcaption}
\usepackage{caption}
\usepackage{xurl}
\usepackage[hidelinks]{hyperref}
\graphicspath{ {./images/} }
\renewcommand{\figurename}{Fig.}

\makeatletter
\newcommand{\linebreakand}{%
  \end{@IEEEauthorhalign}
  \hfill\mbox{}\par
  \mbox{}\hfill\begin{@IEEEauthorhalign}
}
\makeatother

\def\BibTeX{{\rm B\kern-.05em{\sc i\kern-.025em b}\kern-.08em
    T\kern-.1667em\lower.7ex\hbox{E}\kern-.125emX}}
\begin{document}

\title{Optimal Multispectral Imaging using RGB Cameras\\
\thanks{This work has been supported by the Croatian Government project NPOO.C3.2.R3-I1.04.0033 "Scalable System of Cameras and Optical Filters for Industrial Applications (SKORPI)" 2024.}
}

\author{\IEEEauthorblockN{1\textsuperscript{st} Tomislav Matulić}
\IEEEauthorblockA{\textit{Faculty of Electrical Engineering and Computing} \\
\textit{University of Zagreb}\\
Zagreb, Croatia \\
tomislav.matulic@fer.hr}
\and
\IEEEauthorblockN{2\textsuperscript{nd} Ivan Škrabo}
\IEEEauthorblockA{\textit{Faculty of Electrical Engineering and Computing} \\
\textit{University of Zagreb}\\
Zagreb, Croatia \\
ivan.skrabo@fer.hr}
\linebreakand 
\IEEEauthorblockN{3\textsuperscript{rd} Dubravko Babić}
\IEEEauthorblockA{\textit{Faculty of Electrical Engineering and Computing} \\
\textit{University of Zagreb}\\
Zagreb, Croatia \\
dubravko.babic@fer.hr}
\and
\IEEEauthorblockN{4\textsuperscript{th} Damir Seršić}
\IEEEauthorblockA{\textit{Faculty of Electrical Engineering and Computing}} 
\textit{University of Zagreb}\\
Zagreb, Croatia \\
damir.sersic@fer.hr}

\maketitle

\begin{abstract}
We present a physics-driven framework for accurate evaluation of discrete spectral bands using a low-cost multispectral setup built from off-the-shelf RGB cameras and narrow multi-band optical filters. The approach starts by explicitly formulating a linear measurement model. The camera responses are expressed as linear mixtures of unknown spectral components, with mixing coefficients determined by the overlap between the camera spectral sensitivities and the filter transmittances. For a multi-camera configuration, the per-camera models are stacked into a single global system whose structure is fully determined by the allocation of target wavelengths across the camera--filter units. We pose wavelength allocation as a deterministic design problem and select the configuration that minimizes the spectral condition number of the resulting system matrix. Guided by a frame-theoretic interpretation, this criterion promotes numerical stability, maximizes worst-case output signal-to-noise ratio, and improves the robustness of spectral reconstruction. The design space is finite, enabling the evaluation of all feasible configurations under practical constraints. We demonstrate the method on a representative example with 12 target wavelengths and four triband filters, and identify the wavelength allocation that yields the most stable and noise-robust recovery. The proposed framework includes redundant configurations, in which individual wavelengths are measured by multiple cameras, thereby providing additional degrees of freedom that further improve noise robustness.

\end{abstract}


\section{Introduction}

Multispectral imaging (MSI) refers to the acquisition of co-registered images of the same scene in a small number of distinct, discretely positioned spectral bands. By sampling the radiance of the scene at multiple wavelengths, MSI provides information that is inaccessible to standard RGB imaging, enabling improved material discrimination and quantitative spectral analysis. As a result, MSI has become an important tool in remote sensing and agriculture, medical imaging, food safety and quality inspection, cultural-heritage documentation and art conservation~\cite{Zhang2025, Ilianu2023, Su2018, Qin2013, Jones2020}.

Despite these benefits, MSI hardware is often substantially more complex and costly than standard RGB imaging systems. Many acquisition concepts rely on additional wavelength-selection optics and associated calibration procedures, which increase cost and system complexity, and impose trade-offs in acquisition speed and robustness. A practical alternative is therefore to construct MSI systems from mass-produced RGB sensors combined with carefully designed narrow bandpass optical filters. In such configurations, the measured RGB values represent linear mixtures of the unknown spectral components, so spectral recovery becomes an inverse problem whose accuracy depends strongly on the forward model and its numerical conditioning. 

The primary motivation for using off-the-shelf RGB cameras combined with tri-band optical filters is the ability to customize the spectral response of the MSI system after the sensors and cameras have been manufactured. In contrast, many multispectral cameras rely on multi-zone (mosaic) filters deposited directly on the sensor (Bayer patterns), where 16, 25, or more wavelength channels are permanently assigned to pixels during fabrication. Because this wavelength layout is fixed at the manufacturing time, the selection of spectral bands cannot be modified once the camera is built. However, with external tri-band filters, the user can reconfigure the wavelength set simply by replacing the filters mounted in front of the RGB cameras. This post-manufacturing reconfigurability provides a level of flexibility that is typically not available in conventional mosaic-filter multispectral cameras.

The central aim of this work is to enable accurate evaluation of spectral bands in an off-the-shelf multi-camera configuration by (i) explicitly formulating the measurement model and (ii) guiding the filter/camera wavelength allocation toward numerically stable and noise-robust reconstruction.

\subsection{Related work}
Alternative low-cost MSI systems have been demonstrated in several hardware configurations, including multi-camera and filtered-RGB concepts~\cite{Ohsawa2004, Park2007, Shrestha2011}. These systems are naturally described by a linear forward model relating the scene spectral radiance to sensor responses. Recovering spectra from the measurements has therefore been widely studied, ranging from classical linear regression and basis-projection approaches to modern learning-based reconstruction methods~\cite{Imai2000, Cao2024, Huang2022}. While learning-based methods can achieve strong performance, physics-driven linear models remain attractive when calibration transparency, interpretability, and deterministic behavior are required.

A critical factor in linear reconstruction is the conditioning of the system matrix. It maps spectral samples to observations: poorly conditioned systems amplify noise and yield unstable estimates. This has motivated filter selection and band allocation strategies that explicitly promote numerical stability. For example, Hardeberg \emph{et al.}~\cite{Hardeberg2004FilterSF} proposed selecting filters by maximizing mutual orthogonality of response vectors, aiming to span the spectral space broadly, but without directly optimizing inversion stability. Li \emph{et al.}~\cite{Li2018} introduced a maximum linear independence (MLI) criterion that improves invertibility by minimizing the condition number (i.e., the ratio of the largest to the smallest singular value) of the sensitivity matrix. MLI is primarily used as a principled rule for choosing an informative subset of filters from a larger library, aiming to reduce ambiguity among channels and to stabilize subsequent spectral reconstruction. Efficient search procedures for exploring filter combinations under such criteria have also been proposed, such as Sippel's fast binary search strategy~\cite{Sippel2022}. In parallel, multi-camera MSI arrays built from off-the-shelf components have confirmed the practicality of extending spectral sensitivity via multiple filtered cameras~\cite{Genser2020}. A closely related concept was demonstrated by Themelis \emph{et al.}~\cite{Themelis_08}, who used standard color cameras combined with multiple-bandpass filters to obtain a calibrated linear mixing model allowing simultaneous recovery of several spectral band intensities via matrix inversion.

\subsection{Proposed approach}
In contrast to task-specific or heuristic designs, we propose a systematic framework for multi-camera MSI design. We formulate a linear forward model for an arbitrary number of cameras, individual filter passbands, and target wavelengths, and derive analytic stability bounds linking numerical conditioning to worst-case noise amplification. Based on these insights, we enumerate feasible camera/filter configurations and select the wavelength allocation that minimizes the condition number of the global system matrix. This yields a fully deterministic and reproducible design procedure for allocating wavelengths to each camera's filter such that spectral recovery is maximally noise-robust under the given hardware constraints. Whereas Li \emph{et al.}~\cite{Li2018} use the MLI criterion to select an informative filter subset primarily by minimizing the condition number of the filter transmittance matrix, we instead optimize the conditioning of the full camera--filter forward model for a prescribed set of target wavelengths and a general multi-camera wavelength-allocation setting. This shifts the objective from analyzing the filters in isolation to guaranteeing a well-posed, noise-robust inversion for the specific multi-camera configuration considered in this paper.

Throughout the paper, we assume a predefined finite set of target wavelengths at which spectral information is to be recovered. For each camera, the combination of the sensor’s spectral sensitivities and a multi-band optical filter induces mixing coefficients determined by the overlap between the filter transmittances and the camera response functions. This yields a linear system that relates the desired spectral samples to the measured values. For instance, consider $6$ target wavelengths and $2$ RGB cameras, each equipped with a $3$-band filter (so that $3\times 2 = 6$): the design task is to distribute the target wavelengths across the two filters such that the resulting inverse problem is well-conditioned and robust to noise. This paper addresses precisely this allocation problem for a general number of cameras, filter passbands, and target wavelengths, enabling stable and most accurate recovery of the prescribed spectral bands.

\begin{center}
\begin{figure}
    \centering
    \includegraphics[width=0.9\linewidth]{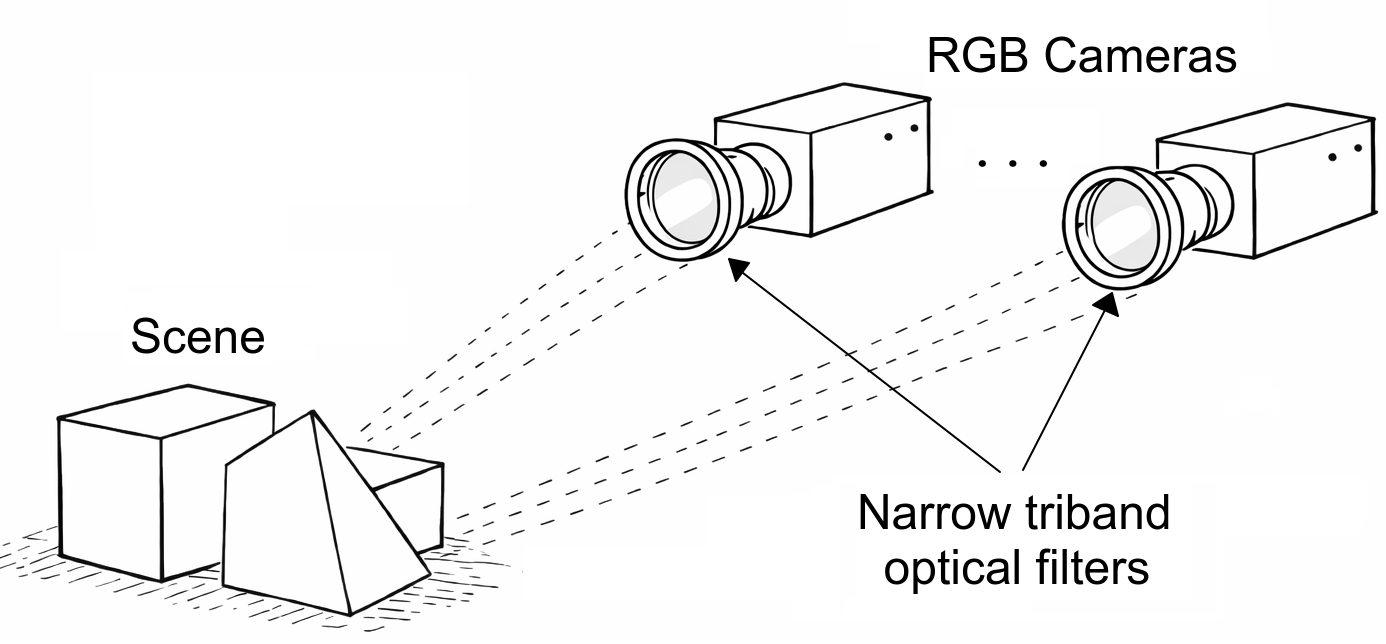}
    \caption{Acquisition setup of the proposed multi-camera MSI system. Multiple off-the-shelf RGB cameras observe the same scene, each equipped with a narrow triband  optical filter. The ensemble of camera/filter pairs provides multiplexed spectral measurements that are used to recover scene radiance at a predefined set of target wavelengths.}
    \label{fig:Setup}
\end{figure}
\end{center}
\subsection{Paper organization}
Section~II develops the theoretical foundation: we derive the linear forward model and interpret the resulting stability bounds through a frame-theoretic perspective, which directly motivates the proposed conditioning-based design criterion. Section~III presents representative design examples(Ivane?). Section~IV concludes the paper and discusses practical implications.

\section{Theoretical framework}\label{sec:2}

Assume a $C$-channel camera with spectral sensitivity functions $S_c(\lambda)$ for $c=1,2,\dots,C $, where $S_c(\lambda)$ includes the sensor responsivity and the transmission of the multizone filter on the sensor. At a single pixel location $(x,y)$, the recorded value in channel $c$, denoted by $y_c~=~y_c(x,y)$, can be described by the linear forward model 
\begin{equation}
    y_c \;=\; \int\limits_{\lambda} E(\lambda)\,S_c(\lambda)\,d\lambda \;+\; n_c.
    \label{eq:model_cam}
\end{equation}
$E(\lambda)$ is the spectral irradiance at the sensor and $n_c$ is an additive noise term that accounts for multiple sources (e.g., photon shot noise due to the discrete nature of light, electronic/readout noise, and additional acquisition/processing noise). In what follows, we assume that $n_c$ is a white noise. In practice, the most common case is $C=3$ (RGB cameras), although the same formulation applies to multispectral sensors with $C>3$ channels.

We begin by considering a single $C$-channel camera equipped with a single $k$-band narrow bandpass optical filter.
Let $F_i(\lambda)$, $i=1,2,\dots,k$, denote the spectral transmittance of the $i$-th bandpass, with central wavelength $\lambda_i$.
When the filter is placed in front of the camera lens, the measurement model in \eqref{eq:model_cam} becomes
\begin{equation}
\begin{aligned}
    y_c
    &= \int_{\lambda} E(\lambda)\Big(\sum_{i=1}^{k} F_i(\lambda)\Big) S_c(\lambda)\,d\lambda \;+\; n_c \\
    &= \sum_{i=1}^{k} \int_{\lambda} E(\lambda)\,F_i(\lambda)\,S_c(\lambda)\,d\lambda \;+\; n_c.
\end{aligned}
\label{eq:model_cam_filter_start}
\end{equation}
We assume that each bandpass is sufficiently narrow so that the spectral irradiance is approximately constant over the support of $F_i(\lambda)$.
We denote the spectral irradiance around $\lambda_i$ by $x_i$ , i.e., $x_i \approx E(\lambda_i)$, and define
\[
    \alpha_{ci} \;=\; \int_{\lambda} F_i(\lambda)\,S_c(\lambda)\,d\lambda
\]
as the effective mixing coefficient that quantifies the overlap between the $i$-th bandpass transmittance $F_i(\lambda)$ and the spectral sensitivity $S_c(\lambda)$ of the channel $c$.
Consequently, we obtain the approximation
\begin{equation}
\begin{aligned}
    y_c
    &\approx \sum_{i=1}^{k} \left(\int_{\lambda} F_i(\lambda)\,S_c(\lambda)\,d\lambda\right) x_i \;+\; n_c \\
    &= \sum_{i=1}^{k} \alpha_{ci}\,x_i \;+\; n_c.
\end{aligned}
\label{eq:model_cam_filter}
\end{equation}

The expression \eqref{eq:model_cam_filter} can be written compactly in matrix form as
\begin{equation}
    \bm y_s \;=\; \bm A_s\,\bm x_s \;+\; \bm n_s,
    \label{eq:model_cam_matrix}
\end{equation}
where 
\begin{align*}
   \bm x_s&=[x_1\,x_2\,\dots\,x_k]^T\in\mathbb{R}^{k},\\
   \bm y_s&=[y_1\,y_2\,\dots\,y_C]^T\in\mathbb{R}^{C}, \\
   \bm A_s&=[\alpha_{ci}]\in\mathbb{R}^{C\times k},\\
   \bm n_s&=[n_1\,n_2\,\dots\,n_C]^T\in\mathbb{R}^{C}.
\end{align*} 
The coefficients $\alpha_{ci}$, and thus the matrix $\bm A_s$, are assumed to be known: the filter spectral responses $F_i(\lambda)$ are determined by the filter design, while the camera spectral sensitivity curves $S_c(\lambda)$ are typically provided by the manufacturer or obtained via calibration. The measurement vector $\bm y_s$ is also known, as it represents the measurements. 

Thus, the matrix $\bm A_s$ defines the linear mapping between the unknown vector $\bm x_s\in\mathbb{R}^k$ whose entries represent the spectral irradiance at the selected $k$ wavelength bands (i.e., $x_i \approx E(\lambda_i)$), and the measured sensor responses $\bm y_s$ recorded by a single $C$-channel camera, up to the additive noise term $\bm n_s$.

\subsection{Multiple cameras and optical filters}

Having established the forward model for a single camera equipped with a $k$-band narrow bandpass filter, we now extend the framework to a multi-camera acquisition system. Let
\[
    K_0 \;=\; \{\lambda_1,\lambda_2,\dots,\lambda_p\}
\]
denote the set of $p$ target central wavelengths at which we wish to acquire the multispectral information. We consider a system of $N_{\mathrm{cam}}$ cameras, where each camera is equipped with a $k$-band filter chosen overall set $K_0$. For each camera $i\in\{1,\dots,N_{\mathrm{cam}}\}$ we assign a $k$-element subset
\[
    K_i \;\subset\; K_0,
    \qquad \mathrm{card}\!\left(K_i\right)= k,
\]
which specifies the central wavelengths of the passbands implemented in the corresponding optical filter.

To ensure that the overall system covers the entire set of desired wavelengths, the total number of available passbands across all cameras must be at least $p$, i.e.,
\[
    k\,N_{\mathrm{cam}} \;\geq\; p.
\]

In what follows, for simplicity, we assume that all cameras are identical (or sufficiently well matched), i.e., they share the same channel-wise spectral sensitivity functions $S_c(\lambda)$ for $c=1,\dots,C$, and have the exposure time. This assumption isolates the effect of wavelength allocation and filter design from inter-camera variability and allows the system matrix to be expressed using a single set of sensitivity curves.

Under this assumption, we introduce the design matrix  $\bm D \in \mathbb{R}^{C \times p}$ with entries
\begin{equation}
    d_{ci} \;=\; \alpha_{ci} \;=\; \int\limits_{\lambda} F_i(\lambda)\,S_c(\lambda)\,d\lambda,
    \label{eq:Design_coefs}
\end{equation}
where $F_i(\lambda)$ denotes the spectral transmittance of the $i$-th narrowband filter centered at $\lambda_i$. Thus, $d_{ci}$ is the effective mixing coefficient that quantifies the contribution of the $i$-th wavelength band to channel $c$; it captures the overlap between the transmittance profile $F_i(\lambda)$ of the $i$-th bandpass filter and the spectral sensitivity $S_c(\lambda)$ of channel $c$. Consequently, the columns of the design matrix
\[
    \bm D \;=\; [\,\bm d_{1}\; \bm d_{2}\; \cdots\; \bm d_{p}\,] \in \mathbb{R}^{C\times p}
\]
represent, for each wavelength band $\lambda_i$, the vector of overlaps between the corresponding filter transmittance $F_i(\lambda)$ and the camera spectral sensitivities $S_c(\lambda)$ across all channels $c=1,\dots,C$.  In practice, the design matrix $\bm D$ is full rank, which follows from the fact that the channel spectral sensitivities $S_c(\lambda)$ are sufficiently distinct.

Importantly, the proposed optimization is not restricted to identical cameras and applies equally to heterogeneous multi-camera systems. In that case, each camera $i$ is characterized by its own set of channel sensitivities $S_{i,c}(\lambda)$, and the corresponding rows of the forward model are constructed using the appropriate camera-specific responses. For example, one may combine an RGBI camera that extends into the near-infrared with a second camera featuring UV-extended responsivity. The wavelength-allocation procedure and the conditioning-based criterion remain unchanged: only the underlying sensitivity functions used to populate the design matrix differ. Consequently, the framework can be used both for matched-camera arrays and for deliberately heterogeneous camera ensembles designed to broaden the overall spectral coverage.

As a first step, we use the matrix $\bm D$ to describe a single camera/optical-filter system by masking $\bm D$:
we retain only the $k$ columns corresponding to the selected central wavelengths $\lambda_i$, while setting all remaining columns to zero.
Subsequently, the resulting per-camera matrices are stacked up to form a single global linear system that jointly models all
camera--filter units in the multi-camera acquisition setup.

Let us proceed step by step: by selecting $k$ columns of the matrix $\bm D$ and setting all remaining columns to zero, i.e., by choosing $k$ wavelength bands for a given optical filter/camera configuration, we obtain
\[
    \bm D_{K} \;=\; [\bm0\;\cdots \;\bm d_{i_1}\; \bm0\;\cdots\;\bm d_{i_2}\; \bm0 \cdots\; \bm d_{i_k}\,\bm0\;\cdots\; \bm 0],
\]
where $K = \{\lambda_{i_1}, \lambda_{i_2}, \dots, \lambda_{i_k}\}$ denotes the selected subset of wavelengths. In other words, $\bm D_{K}$ acts as a column mask on $\bm D$: it retains only the columns associated with wavelengths in $K$ and sets all remaining columns to the zero vector, thereby nulling their contribution to the measurement model. In this way, each choice of $k$ distinct wavelengths corresponds to one specific filter configuration. Since the ordering within the subset is irrelevant, the total number of distinct configurations that can be formed from $p$ candidate wavelengths is
\[
    \binom{p}{k},
\]
each yielding a different matrix $\bm D_{K}$.

Using the matrix $\bm D_K$, the forward model for a single camera/optical filter can be written as
\begin{equation}
    \bm y_K \;=\; \bm D_K\,\bm x \;+\; \bm n_K,
    \label{eq:model_cam_proj_D}
\end{equation}
where
\begin{align*}
   \bm x &=[x_1\,x_2\,\dots\,x_p]^T \in \mathbb{R}^{p},\\
   \bm y_K &=[y_1\,y_2\,\dots\,y_C]^T \in \mathbb{R}^{C}, \\
   \bm n_K &=[n_1\,n_2\,\dots\,n_C]^T \in \mathbb{R}^{C}.
\end{align*}
The components of $\bm x$ represent the spectral irradiance sampled at the candidate wavelengths, i.e.,
$x_i \approx E(\lambda_i)$ for $i=1,2,\dots,p$, while $\bm y_K$ and $\bm n_K$ denote the corresponding measured camera responses and the associated noise, respectively.

We now focus on the multi-camera setup: we have a total of $N_{\mathrm{cam}}$ subsets of $K_0$, each containing exactly $k$ elements, one for each camera. 
Therefore, for each camera $i=1,\dots,N_{\mathrm{cam}}$, we select a $k$-element subset $K_i \subset K_0$, where $K_i$ specifies the $k$ central wavelengths used by the corresponding narrowband filter. 
For each subset $K_i$, we construct the matrix $\bm D_{K_i}$ as described previously. A full multi-camera design is therefore determined by a configuration
\[
    \mathcal{C}_j \;=\; \{K_1, K_2, \dots, K_{N_{\mathrm{cam}}}\}.
\]
For each camera $i$ and corresponding matrix $\bm D_{K_i}\in\mathbb{R}^{C\times p}$, the per-camera measurement model is
\[
    \bm y_{K_i} \;=\; \bm D_{K_i}\,\bm x \;+\; \bm n_{K_i},
    \qquad i=1,\dots,N_{\mathrm{cam}}.
\]
By concatenating (stacking) the measurements from all cameras, the full multi-camera system can be written as
\begin{equation}
    \bm y \;=\; \bm A_{\mathcal{C}_j}\,\bm x \;+\; \bm n,
    \label{eq:block_system}
\end{equation}
where the global system matrix is given by
\begin{equation}
\begin{aligned}
    \bm A_{\mathcal{C}_j} = 
    \begin{bmatrix}
        \bm D_{K_1} \\
        \bm D_{K_2} \\
        \vdots \\
        \bm D_{K_{N_{\mathrm{cam}}}}
    \end{bmatrix}
    \in \mathbb{R}^{\left(N_{\mathrm{cam}}C\right)\times p}.
\end{aligned}
\end{equation}
The stacked measurement and noise vectors are defined as
\begin{align*}
    \bm y
    &=
    \big[\bm y_{K_1}^\top\ \bm y_{K_2}^\top\ \cdots\ \bm y_{K_{N_{\mathrm{cam}}}}^\top\big]^\top
    \in \mathbb{R}^{N_{\mathrm{cam}}C},\\
    \bm n
    &=
    \big[\bm n_{K_1}^\top\ \bm n_{K_2}^\top\ \cdots\ \bm n_{K_{N_{\mathrm{cam}}}}^\top\big]^\top
    \in \mathbb{R}^{N_{\mathrm{cam}}C}.
\end{align*}
The vector $\bm y$ collects all measurements acquired across the $N_{\mathrm{cam}}$ cameras, each observing the scene through its corresponding $k$-band optical filter.

Finally, we emphasize that the selected collection of wavelength subsets
$\{K_i\}_{i=1}^{N_{\mathrm{cam}}}$ must satisfy the following feasibility conditions:
\begin{enumerate}
    \item[(i)] \textbf{Non-redundancy:}\quad
    $K_i \neq K_j$ for all $i\neq j$,
    which prevents two cameras from using identical wavelength sets and thus avoids purely redundant measurements.

    \item[(ii)] \textbf{Full column-rank:}\quad
    The sensing matrix $\bm A_{\mathcal C_j}$ must be full column-rank.
    This condition ensures that the information associated with each selected central wavelength is preserved and that the corresponding inverse problem admits a unique least-squares estimate.

    \item[(iii)] \textbf{Determined or overdetermined system:}\quad
    The overall system must satisfy $N_{\mathrm{cam}}\,C \ge p$,
    ensuring that the number of available measurements is sufficient to recover the $p$ target wavelength components.
\end{enumerate}
Note that these conditions also satisfy the coverage requirement
\[
\bigcup_{i=1}^{N_{\mathrm{cam}}} K_i \;=\; K_0.
\]
Altogether, this guarantees that the resulting matrix $\bm A_{\mathcal C_j}$ is well defined and that every target wavelength in $K_0$ is measured by at least one camera/filter pair. We denote the set of all such feasible subsets by $\mathcal{C}$.

\subsection{Minimum case}
In the special case $k=C$ and $C N_{\mathrm{cam}} = p$, the interpretation simplifies significantly. The system matrix $\bm A_{\mathcal{C}_j}$ is square, and each target wavelength is represented exactly once across camera/filter pair (i.e., there is no repetition). This corresponds to the minimum acquisition setting, in the sense that it uses the smallest possible number of cameras to cover all $p$ wavelengths.

We first note the following: due to the zero columns in $\bm D_K$, Eq.~\eqref{eq:model_cam_proj_D} can be reduced to the form in Eq.~\eqref{eq:model_cam_matrix} by restricting the model to the selected wavelengths in $K$. In particular, one obtains the reduced matrix
\begin{equation}
\bm A_s \;=\; \bm A_{s,K} \;=\; \big[\bm d_{i_1}\; \bm d_{i_2}\; \cdots\; \bm d_{i_k}\big],
 \label{eq:Design_matrix_reduced}   
\end{equation}
and the corresponding vector $\bm x_s$ contains only the unknown irradiance values associated with the wavelengths in $K$.

In this special setting, $\bm A_{\mathcal{C}_j}$ can be viewed as a column-permuted block-diagonal matrix assembled from the per-camera reduced design matrices introduced in Eq.~\eqref{eq:model_cam_matrix} and Eq.~\eqref{eq:Design_matrix_reduced}. Specifically,
\[
    \bm A_{\mathcal{C}_j} \;=\; \bm B_{\mathcal{C}_j}\,\bm P,
\]
where $\bm P\in\mathbb{R}^{p\times p}$ is a permutation matrix (reordering columns according to the allocation $\mathcal{C}_j$), and
\[
\begin{aligned}
    \bm B_{\mathcal{C}_j}
    \;&=\;
    \mathrm{blkdiag}\!\big(\bm A_{s,K_1},\bm A_{s,K_2},\dots,\bm A_{s,K_{N_{\mathrm{cam}}}}\big)\\
    \;&=\;
    \begin{bmatrix}
        \bm A_{s,K_1} & \bm 0        & \cdots & \bm 0 \\
        \bm 0         & \bm A_{s,K_2} & \cdots & \bm 0 \\
        \vdots        & \vdots        & \ddots & \vdots \\
        \bm 0         & \bm 0         & \cdots & \bm A_{s,K_{N_{\mathrm{cam}}}}
    \end{bmatrix}
    \in \mathbb{R}^{p\times p}.
\end{aligned}
\]
Note that this particular configuration is especially convenient, as it enables us to analyze (and solve problem for) each camera independently. Since $\bm B_{\mathcal{C}_j}$ is block-diagonal, the corresponding system decouples into $N_{\mathrm{cam}}$ separate subproblems, and the solution components of $\bm x$ associated with a given camera are linked only to the corresponding block. As a result, both the forward equations and their inverses are fully separable across cameras. Moreover, $\bm B_{\mathcal{C}_j}$ is invertible, and each diagonal block $\bm A_{s,K_i}\in\mathbb{R}^{C\times C}$ is itself square and invertible, which ensures that each subproblem is well-posed. The permutation matrix $\bm P$ can be interpreted as a reindexing operator that permutes the entries of $\bm x$ such that the wavelength components are ordered consistently with the camera-wise allocation $\mathcal{C}_j$.

\subsection{Inverse Problem and Numerical Properties}

Having fully specified the forward model of the proposed optical acquisition system, we now turn to the inverse problem - recovering the spectral irradiance vector from the multi-camera measurements. In addition, we identify the key numerical properties of the resulting linear system, which will later be used to formulate a selection criterion to identify the most suitable configuration among all feasible choices in $\mathcal{C}$.

Specifically, we consider the linear model from Eq.~\eqref{eq:block_system}. In the case where $N_{\mathrm{cam}}C \geq p$, the system is determined or overdetermined, and we may seek an estimate of $\bm x$ that minimizes the squared residual. This leads to the standard least-squares (LS) solution.
This yields the normal equations (for notational simplicity, we use the shorthand $\bm A = \bm A_{\mathcal{C}_j}$):
\begin{equation}
    \bm A^\top \bm A\, \bm x \;=\; \bm A^\top \bm y.
    \label{eq:LS_T}
\end{equation}
If the matrix $\bm A^\top \bm A$ is full rank, i.e., if $\mathrm{rank}(\bm A) = p$, the least-squares solution is unique and given by
\begin{equation}
    \bm x_{\mathrm{LS}} 
    \;=\; \bigl(\bm A^\top \bm A\bigr)^{-1} \bm A^\top \bm y
    \;=\; \bm A^+ \bm y.
    \label{eq:LS}
\end{equation}
Here, $\bm A^+$ denotes the pseudoinverse, which in this full-column-rank case coincides with
\[
    \bm A^+ \;=\; \bigl(\bm A^\top \bm A\bigr)^{-1} \bm A^\top.
\]
This solution is optimal in the sense that it minimizes the squared $\ell_2$-norm of the residual (noise term)
\[
    \|\bm n\|_2^2 
    \;=\; \|\bm y - \bm A \bm x\|_2^2,
\]
over all possible $\bm x$. Moreover, if the noise components are independent and Gaussian, i.e.,
$
\bm n \sim \mathcal{N}(\bm 0,\sigma^2 \bm I),
$
then the least-squares estimate coincides with the maximum-likelihood (ML) estimate. 

In the already discussed special case $C=k$ and $C N_{\mathrm{cam}}~=~p$, the system matrix $\bm A$ is square and invertible. Consequently, the Moore--Penrose pseudoinverse reduces to the usual matrix inverse, i.e., $\bm A^{+}=\bm A^{-1}$, and we recover the (unique) solution
\[
    \tilde{\bm x} \;=\; \bm A^{-1}\bm y.
\]
In contrast, if $C N_{\mathrm{cam}}<p$, the stacked system becomes underdetermined and, in general, ill-posed, since there exist infinitely many vectors $\bm x$ that fit the measurements equally well. This regime is excluded by our feasibility constraints on the wavelength allocation, which are chosen to ensure that the overall system admits a well-defined reconstruction of the desired unknowns.

Of particular interest is how the linear operator $\bm A$ acts on a vector $\bm x$, i.e., how it stretches or contracts the vector and, consequently, how the energy of $\bm x$ changes under the transformation induced by $\bm A$. For a real matrix $\bm A$, this behavior is conveniently characterized by the symmetric matrix
\[
    \bm A^\top \bm A,
\]
which encodes how the squared norm of $\bm x$ is mapped to the squared norm of $\bm A\bm x$. The matrix $\bm A^\top \bm A$ is real, square, symmetric, and positive semidefinite, and it already appears in the normal equation from Eq.~\eqref{eq:LS_T}. 

We are interested in how the energy of a vector $\bm x$ changes after applying $\bm A$, that is, we seek positive constants $C_1$ and $C_2$ such that
\begin{equation}
    C_1 \,\|\bm x\|_2^2 \;\leq\; \|\bm A \bm x \|_2^2 \;\leq\; C_2 \,\|\bm x\|_2^2
    \quad\text{for all } \bm x .
    \label{eq:frames}
\end{equation}
The results from frame theory tell us that the optimal (tightest) constants are given by the extremal eigenvalues of $\bm A^\top \bm A$: the lower frame bound $C_1$ equals the smallest eigenvalue $s_{\min}(\bm A^\top \bm A)$, while the upper frame bound $C_2$ equals its largest eigenvalue $s_{\max}(\bm A^\top \bm A)$. Note that the eigenvalues of $\bm A^\top \bm A$ are real, nonnegative, and strictly positive when $\bm A^\top \bm A$ is full rank (i.e., when $\bm A$ has full column rank).

Consequently, the norm of any vector $\bm x$ after transformation $\bm A$ is scaled by a factor lying between $s_{\min}(\bm A^\top \bm A)$ and $s_{\max}(\bm A^\top \bm A)$, and the corresponding energy $\|\bm A \bm x\|_2^2$ is restricted to this range. In the special case $s_{\min}(\bm A^\top \bm A) = s_{\max}(\bm A^\top \bm A) = e_0$ (all eigenvalues of $\bm A^\top \bm A$ are equal), the mapping is a scaled isometry: $\|\bm A \bm x\|_2 = e_0\|\bm x\|_2$ for all $\bm x$. Such systems are called tight frames, and in this case a Parseval-type identity holds for the associated transform.

The ratio of the largest $\sigma_{\mathrm{max}}(\bm A)$ and the smallest $\sigma_{\mathrm{min}}(\bm A)$ singular value,
\begin{equation}
    \kappa(\bm A) \;=\; \frac{\sigma_{\max}(\bm A)}{\sigma_{\min}(\bm A)} \;=\; \sqrt{\frac{s_{\max}(\bm A^\top \bm A)}{s_{\min}(\bm A^\top \bm A)}},
    \label{eq:kappa}
\end{equation}
is known as the spectral condition number of $\bm A$.  It satisfies $\kappa(\bm A) \geq 1$, with $\kappa(\bm A) = 1$ precisely in the tight-frame (energy-preserving) case. The larger the condition number, the more anisotropic the action of $\bm A$ on different directions in the space, and the more energy can be distorted by transformation. In numerical terms, a large condition number indicates an ill-conditioned problem, where small perturbations in the data $\bm y$ may lead to large errors in the reconstructed solution $\bm x$.

\subsection{Design criterion}

For each feasible multi-camera configuration $\mathcal{C}_j~\!=~\!\{K_1,\dots,K_{N_{\mathrm{cam}}}\}\in\mathcal{C}$, we construct the corresponding system matrix $\bm A_{\mathcal{C}_j}$ in \eqref{eq:block_system}, as described in the previous subsections.
To quantify the numerical stability of the resulting inverse problem, we evaluate the spectral condition number of $\bm A_{\mathcal{C}_j}$ (cf.~Eq.~\eqref{eq:kappa}),
\begin{equation}
    \kappa\!\left(\bm A_{\mathcal{C}_j}\right)
    \;=\;
    \frac{\sigma_{\max}\!\left(\bm A_{\mathcal{C}_j}\right)}
         {\sigma_{\min}\!\left(\bm A_{\mathcal{C}_j}\right)} ,
\end{equation}
and adopt the design criterion
\begin{equation}
    \min_{\mathcal{C}_j\in\mathcal{C}} \ \kappa\!\left(\bm A_{\mathcal{C}_j}\right).
    \label{eq:minimize}
\end{equation}

Beyond the advantages and properties discussed above, the condition number $\kappa(\bm A)$ admits a useful worst-case
signal-to-noise interpretation. In particular, by the frame bounds in \eqref{eq:frames}, the output signal energy is
attenuated most severely when the input signal $\bm x$ aligns with the eigenvector corresponding to the smallest eigenvalue
$s_{\min}(\bm A^\top \bm A)$, while the impact of additive noise can be amplified most strongly when the noise component aligns
with the eigenvector associated with the largest eigenvalue $s_{\max}(\bm A^\top \bm A)$. This combination yields the worst-case
output signal-to-noise ratio (SNR). In other words, the output SNR can deteriorate, in the worst case, by a factor
\[
    \frac{s_{\min}(\bm A^\top\bm A)}{s_{\max}(\bm A^\top\bm A)}
    \;=\; \frac{1}{\kappa(\bm A)^2}.
\]
Therefore, minimizing the criterion in Eq.~\eqref{eq:minimize} simultaneously maximizes the worst-case SNR, yielding configurations that are both better conditioned and more robust to noise.

Thus, the closer $\kappa(\bm A_{\mathcal{C}_j})$ is to $1$, the better conditioned the transformation is: the signal energy is least spread, and the influence of noise on the accuracy of the estimated $\bm x$ is reduced even in unfavorable scenarios. This observation provides a convenient design criterion: among all admissible filter configurations, those yielding smaller values of $\kappa(\bm A_{\mathcal{C}_j})$ are preferable, as they lead to numerically more stable and noise-robust reconstructions.

Finally, note that the search space $\mathcal{C}$ is finite: it consists of all feasible permutations of
$N_{\mathrm{cam}}$ distinct $k$-element subsets whose union covers $K_0$. Therefore, the optimization problem is discrete
(combinatorial) and can, in principle, be solved by exhaustive evaluation and sorting of all feasible configurations.
This yields a conceptually simple and fully deterministic procedure for selecting the best filter arrangement under the
proposed design criterion. In the practical settings considered in this work, the number of feasible configurations remains
moderate (typically on the order of $\times 10^{4}$), making the exhaustive search straightforward on a standard
computer.

\section{Results}

The evaluation was performed using twelve wavelengths: 
\begin{equation}
\begin{aligned}
K_0  = &\{\lambda_1,\lambda_2,\ldots,\lambda_{12}\} \\ 
=&\{410, 430, 450, 500, 520, 550,\\
& \,\,\, 578, 620, 680, 700, 720, 780\}~\mathrm{nm}.
\end{aligned}
\end{equation}
Narrow filters had Gaussian response with full width at half maximum (FWHM) of $10~\mathrm{nm}$. Spectral sensitivity of the camera $S_c(\lambda)$ was determined by the Sony IMX411 sensor properties. The wavelengths set $K_0$ is selected for use in the food industry. Although the Sony sensor is used as a representative example, the proposed framework is sensor-agnostic and applies to any camera, provided that its channel-wise spectral sensitivity is known. Given these specifics and considering that we are using triband ($k = 3$) filters, the minimal number of sensors and triband filters required to cover all $p=12$ wavelengths is $N_{\mathrm{cam}} =4$.

\begin{figure}[h]
    \centering
    \includegraphics[width=0.5\textwidth]{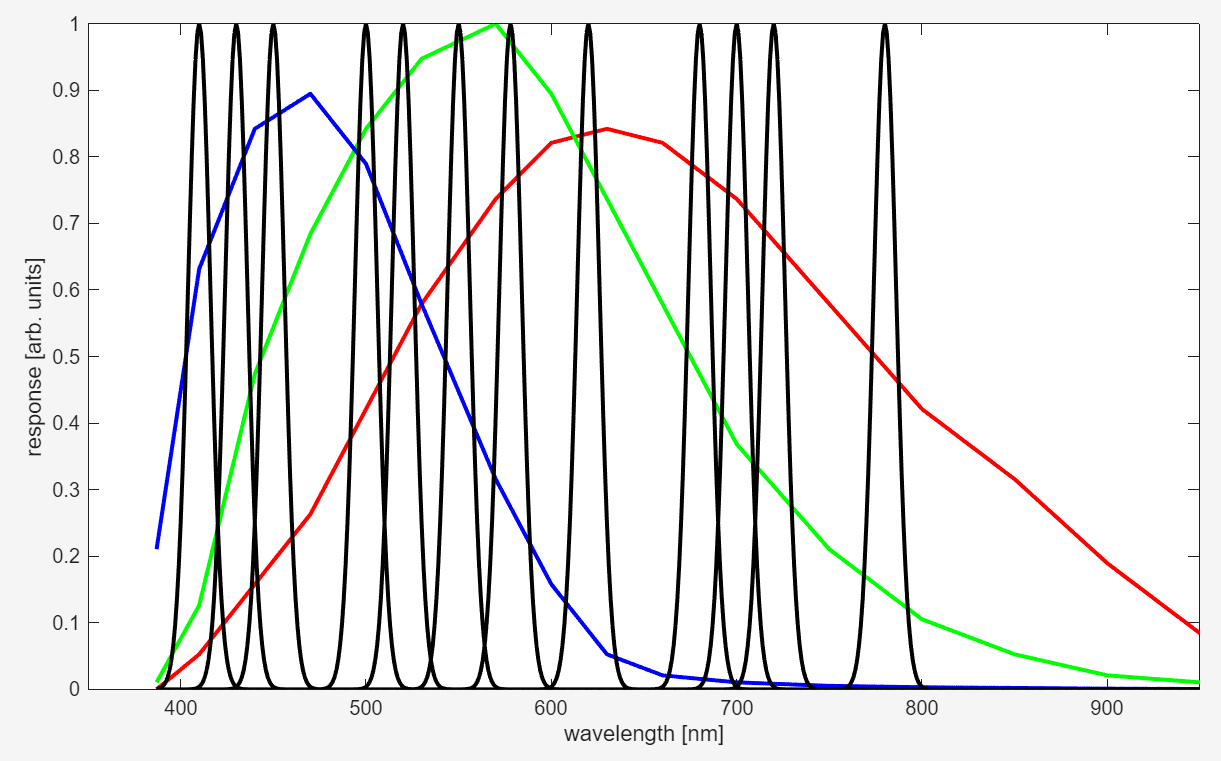}
    \caption{Spectral performance of the CMOS sensor Sony IMX411 used in this work and the assumed set of 12 bandpass filters used in this work as an illustration.}
    \label{spectral_performance}
\end{figure}

Using the sensor's spectral sensitivity data, whose range is from $387~\mathrm{nm}$ to $950~\mathrm{nm}$, we first normalize each channel response by the largest peak among the three channels. Specifically, letting $S_c(\lambda)$, $c\in\{R,G,B\}$ denote the raw spectral sensitivities, we define the normalization factor as
$
1/\max\left(\max(\mathrm{R}),\max(\mathrm{G}),\max(\mathrm{B})\right).
$
Knowing the sensor’s channel-wise spectral sensitivity curves $S_c(\lambda)$ over the full operating range, we approximate each of the 12 target passbands by a Gaussian transmittance function $F_i(\lambda)$ centered at the corresponding wavelength. This allows us to predict the sensor’s effective response at each selected wavelength through the overlap between $S_c(\lambda)$ and $F_i(\lambda)$. The resulting spectral sensitivities together with the 12 Gaussian passbands are shown in Fig.~\ref{spectral_performance}. Consequently, for the known camera response $S_c(\lambda)$ and the prescribed filter transmittances $F_i(\lambda)$, the mixing coefficients $\alpha_{ci}$ in \eqref{eq:Design_coefs} are fully determined.

This enables us to construct the design matrix $\bm D$. In this setup $\bm D$ has size $3\times 12$: each row corresponds to one camera channel, and each column corresponds to one target wavelength in $K_0$. As described in the previous section, $\bm D$ serves as a template from which we construct the system matrices $\bm A_{\mathcal{C}_j}$ for all feasible wavelength allocations $\mathcal{C}_j\in\mathcal{C}$.

All feasible configurations can be counted as
\begin{equation}
\mathrm{card}\!\left(\mathcal{C}\right)
\;=\;
\sum_{i = 0}^{p} (-1)^i\binom{p}{i}\binom{\binom{p - i}{k}}{N_\mathrm{cam}},
\label{eq:cardC_a}
\end{equation}
which follows directly from the inclusion--exclusion principle. The term indexed by $i$ accounts for configurations that omit (i.e., do not measure) at least $i$ out of the $p$ target wavelengths: once $i$ wavelengths are excluded, there remain $p-i$ wavelengths from which one can form \(\binom{p-i}{k}\) distinct $k$-element subsets (candidate $k$-band filters), and \(\binom{\binom{p-i}{k}}{N_\mathrm{cam}}\) counts the choice of $N_\mathrm{cam}$ such subsets (one per camera). Note that the sum is effectively truncated, because terms with \(\binom{p-i}{k}<N_\mathrm{cam}\) vanish (the outer binomial coefficient equals zero). 
For the minimum case, the above expression can be simplified to
\begin{equation}
\begin{aligned}
\mathrm{card}\left(\mathcal{C}\right)
&\;=\;
\frac{\binom{p}{k}\binom{p-k}{k}\binom{p-2k}{k}\cdots\binom{p-(N_{\mathrm{cam}}-1)k}{k}}{N_{\mathrm{cam}}!}\\
&\;=\;\frac{p!}{N_{\mathrm{cam}}!\cdot k!^{N_{\mathrm{cam}}}}
\end{aligned}
\label{eq:cardC_b}
\end{equation}
In our example, this yields \(\mathrm{card}\!\left(\mathcal{C}\right)=15\,400\).

For implementation convenience, we precomputed all feasible allocations and stored them as an index table
\[
\bm I_p \in \{1,\dots,p\}^{15\,400\times 12},
\]
where each row encodes one configuration $\mathcal{C}_j$: the first $k$ entries specify the wavelength indices assigned to the first triband filter, the next $k$ entries correspond to the second filter, and so on. Thus, retrieving the $j$-th row of $\bm I_p$ immediately yields one feasible configuration $\mathcal{C}_j\in\mathcal{C}$. This allows us to construct the corresponding matrix $\bm A_{\mathcal{C}_j}$ without recomputing combinations. In practice, $\bm I_p$ is generated once, saved to disk, and loaded as needed during the evaluation of all configurations.

For each feasible permutation of wavelengths, we get $12 \times 12$ system matrix $\bm A_{\mathcal{C}_j}$. The matrix columns represent which wavelengths are being used and the $12$ rows are representations of the triband filters on cameras. Since each camera filter passes only three wavelengths, each row contains exactly three non-zero entries. As a result, in the special square case, $\bm A_{\mathcal{C}_j}$ can be viewed as block-diagonal up to a column permutation (cf.\ the special square-case discussion).

\begin{figure}[h]
    \centering
    \includegraphics[width=0.5\textwidth]{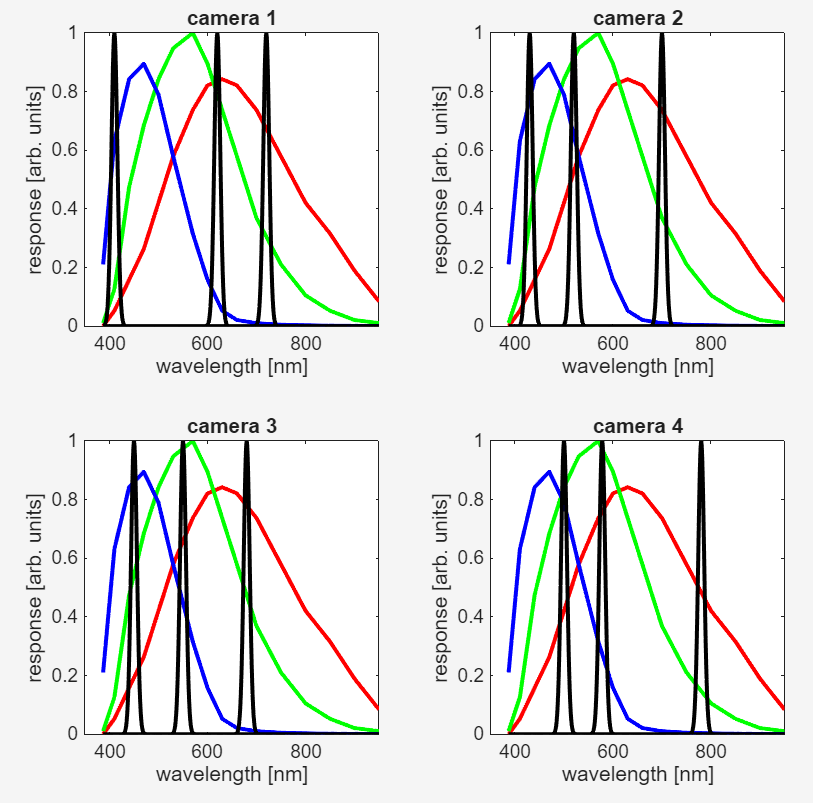}
    \caption{Optimal spectral distribution of the wavelengths onto four narrow triband filters}
    \label{spectral_distribution}
\end{figure}

After constructing the $12\times 12$ matrix for a given allocation, we compute its singular values and evaluate the corresponding spectral condition number (cf.\ \eqref{eq:kappa}). The resulting $\kappa\!\left(\bm A_{\mathcal{C}_j}\right)$ is stored in an array containing all $15{,}400$ feasible configurations. After evaluating all allocations, we select the minimum value, $\kappa_{\min}=9.0840$, which yields the optimal distribution of wavelengths across the four triband filters:
\[
\begin{aligned}
\{&\{410, 620, 720\},
\{430, 520, 700\},\\
&\{450, 550, 680\},
\{500, 578, 780\}\}~\mathrm{nm}. 
\end{aligned}
\]
The distribution is shown in Fig.~\ref{spectral_distribution}. Under the assumed RGB camera model and the above parameters, this configuration is optimal in the sense of noise robustness: it minimizes the condition number of the sensing matrix, thereby maximizing the worst-case output SNR, while also yielding the least spread of signal energy across the transform.

The realization in MATLAB can be found at \cite{implementacija}.

\subsection{Discussion: Use cases}
So far, we have described the complete design procedure for the minimum setting
\(
N_{\mathrm{cam}}C = N_{\mathrm{cam}}k = p,
\)
where the number of measurements matches the number of unknown spectral coefficients and each target wavelength appears exactly once across the camera/filter units. In practice, however, two common situations lead to non-minimal (redundant) designs with
\(
N_{\mathrm{cam}}C > p.
\)
First, the desired number of wavelengths $p$ may not be a multiple of the number of camera channels $C$, making exact one-to-one coverage impossible without repetitions. Second, even when $p$ is compatible with $C$ and $k$, one may wish to improve the reconstruction robustness (SNR) of specific wavelengths by adding cameras, which inevitably introduces repeated measurements of some spectral components. Below we outline these two scenarios using illustrative examples.

\subsubsection{Use case 1: $p$ is not a multiple of the camera channels}
Assume that we wish to estimate one fewer wavelength, i.e., $p=11$, while keeping $N_{\mathrm{cam}}=4$ and $k=C=3$ unchanged. Since the system still provides $N_{\mathrm{cam}}C = 12$ measurements, exactly one wavelength must be repeated across different camera/filter pairs. The design matrix becomes $\bm D\in\mathbb{R}^{3\times 11}$, with entries computed as in~\eqref{eq:Design_coefs}. For each of the $69\,300$ feasible allocations (Eq.~\eqref{eq:cardC_a}), we construct the corresponding system matrix $\bm A_{\mathcal{C}_j}$ by stacking and masking $\bm D$ according to the selected wavelength assignment. In this case,
\(
\bm A_{\mathcal{C}_j}\in\mathbb{R}^{12\times 11},
\)
since it maps the $12$ measurements to the $11$ spectral coefficients to be estimated. The measurement noise therefore has $12$ components, assumed mutually uncorrelated, corresponding to one independent noise term for each camera channel. For every candidate $\bm A_{\mathcal{C}_j}$ we compute the spectral condition number and select the configuration that minimizes it. As discussed previously, a smaller condition number (for a generally rectangular matrix) implies a more stable least-squares inversion and reduced worst-case SNR, thus yielding the desired optimal wavelength allocation.

\subsubsection{Use case 2: Adding additional cameras to improve robustness}
Next, consider adding one more camera/filter pair, i.e., $N_{\mathrm{cam}}=5$, while keeping $p=12$ and $k=3$ unchanged. The design matrix $\bm D\in\mathbb{R}^{3\times 12}$ remains the same, but the global system becomes overdetermined with
\(
N_{\mathrm{cam}}C = 15
\)
measurements, and some wavelengths must be repeated across the five filters. For each of the $32\,501\,700$ feasible allocations (Eq.~\eqref{eq:cardC_a}), we construct the corresponding system matrix
\(
\bm A_{\mathcal{C}_j}\in\mathbb{R}^{15\times 12},
\)
compute its condition number, and choose the configuration with the minimum value. This produces the most numerically stable design under the given hardware constraints, i.e., the allocation with the best worst-case SNR.

\subsubsection{General remarks on redundancy}
More generally, the same workflow applies whenever $N_{\mathrm{cam}}$, $k$, or $p$ changes: one first constructs the design matrix $\bm D$ from the known camera sensitivities and optical filter specifications, then forms $\bm A_{\mathcal{C}_j}$ for each feasible allocation, determines $\kappa(\bm A_{\mathcal{C}_j})$), and finally selects the minimizer over all feasible configurations.

Note that a redundant wavelength allocation can no longer be written in a block-diagonal form (even up to a column permutation); this convenient structure arises only in the non-redundant minimum case, where each target wavelength is assigned exactly once across the camera/filter pairs. Importantly, redundancy can be beneficial: repeating a wavelength provides additional independent measurements of the same spectral coefficient, increasing the effective information available for that component, and often improving the conditioning of the inverse problem. In other words, redundancy typically improves the output SNR.

\section{Conclusion}

This paper considered the design of a cost-effective multispectral acquisition system constructed from off-the-shelf RGB cameras equipped with narrow multi-band optical filters. Starting from a physically motivated linear image formation model, we showed that the overlap between the camera channel sensitivities and the filter transmittances induces mixing coefficients that fully determine the entries of the forward operator. By extending this model to a multi-camera setting, we obtained a stacked linear system, where each feasible configuration corresponds to a particular allocation of the target wavelengths across the camera--filter units.

A central outcome of the theoretical analysis is that the numerical stability of the spectral recovery is governed by the spectral properties of the system matrix. In particular, a frame-theoretic interpretation links the extremal eigenvalues to tight energy-dispersion bounds, while the spectral condition number quantifies worst-case output SNR.
Consequently, minimizing the spectral condition number yields the most noise-robust configuration among the feasible candidates: it limits the error in the least-squares inversion and maximizes the worst-case output SNR, while keeping the spread of signal energy induced by the transform as small as possible, thereby enabling accurate evaluation of the spectral bands.
Moreover, increasing redundancy by adding additional cameras provides additional degrees of freedom that further improves conditioning and enhances noise robustness, subject to the feasibility constraints.

In the presented example, we consider $12$ target wavelengths distributed across $4$ cameras, each equipped with a $3$-band filter. The proposed method enables a practical evaluation of all feasible allocations and identifies the configuration with the smallest condition number, thereby maximizing numerical stability under the assumed hardware constraints. Importantly, the framework is sensor-agnostic: it applies to any camera (not necessarily RGB) as long as its channel-wise spectral sensitivity curves are known.

Future work will focus on experimental validation with real filters and real scenes and extending the formulation to non-identical cameras. An additional extension is to jointly optimize not only the wavelength allocation, but also practical filter parameters such as bandwidths and passband shapes under manufacturability constraints.

\section{Acknowledgment}
This work has been supported by the Croatian Government project NPOO.C3.2.R3-I1.04.0033 "Scalable System of Cameras and Optical Filters for Industrial Applications (SKORPI)" 2024.

\bibliographystyle{IEEEtran}
\bibliography{refs.bib}

@misc{implementacija,
title = {Implementation - Optimal Multispectral Imaging Using RGB Cameras},
author = {Škrabo, Ivan and Matulić, Tomislav and Babić, Drubravo and Seršić, Damir},
year = {2026.},
url = {https://www.zesoi.fer.hr/\_download/repository/OptimalMultispectralImagingUsingRGBCameras.zip}
}

@article{Cao2024,
  title = {Unsupervised spectral reconstruction from RGB images under two lighting conditions},
  volume = {49},
  ISSN = {1539-4794},
  url = {http://dx.doi.org/10.1364/OL.517007},
  DOI = {10.1364/ol.517007},
  number = {8},
  journal = {Optics Letters},
  publisher = {Optica Publishing Group},
  author = {Cao,  Xuheng and Lian,  Yusheng and Liu,  Zilong and Li,  Jin and Wang,  Kaixuan},
  year = {2024},
  month = apr,
  pages = {1993}
}

@inproceedings{Sippel2022,
  title = {Optimal Filter Selection for Multispectral Object Classification Using Fast Binary Search},
  url = {http://dx.doi.org/10.1109/MMSP55362.2022.9949059},
  DOI = {10.1109/mmsp55362.2022.9949059},
  booktitle = {2022 IEEE 24th International Workshop on Multimedia Signal Processing (MMSP)},
  publisher = {IEEE},
  author = {Sippel,  Frank and Seiler,  Jurgen and Kaup,  Andre},
  year = {2022},
  month = sep,
  pages = {1–5}
}

@inproceedings{Hardeberg2004FilterSF,
  title={Filter Selection for Multispectral Color Image Acquisition},
  author={Jon Yngve Hardeberg},
  booktitle={Image Processing, Image Quality, Image Capture Systems Conference},
  year={2004},
  url={https://api.semanticscholar.org/CorpusID:32623590}
}

@article{Themelis_08,
author = {George Themelis and Jung Sun Yoo and Vasilis Ntziachristos},
journal = {Opt. Lett.},
number = {9},
pages = {1023--1025},
publisher = {Optica Publishing Group},
title = {Multispectral imaging using multiple-bandpass filters},
volume = {33},
month = {May},
year = {2008},
url = {https://opg.optica.org/ol/abstract.cfm?URI=ol-33-9-1023},
doi = {10.1364/OL.33.001023},
}

@article{Li2018,
  title = {Filter Selection for Optimizing the Spectral Sensitivity of Broadband Multispectral Cameras Based on Maximum Linear Independence},
  volume = {18},
  ISSN = {1424-8220},
  url = {http://dx.doi.org/10.3390/s18051455},
  DOI = {10.3390/s18051455},
  number = {5},
  journal = {Sensors},
  publisher = {MDPI AG},
  author = {Li,  Sui-Xian},
  year = {2018},
  month = may,
  pages = {1455}
}

@article{Genser2020,
  title = {Camera Array for Multi-Spectral Imaging},
  volume = {29},
  ISSN = {1941-0042},
  url = {http://dx.doi.org/10.1109/TIP.2020.3024738},
  DOI = {10.1109/tip.2020.3024738},
  journal = {IEEE Transactions on Image Processing},
  publisher = {Institute of Electrical and Electronics Engineers (IEEE)},
  author = {Genser,  Nils and Seiler,  Jurgen and Kaup,  Andre},
  year = {2020},
  pages = {9234–9249}
}

@article{Shrestha2011,
  title = {Multispectral imaging using a stereo camera: concept,  design and assessment},
  volume = {2011},
  ISSN = {1687-6180},
  url = {http://dx.doi.org/10.1186/1687-6180-2011-57},
  DOI = {10.1186/1687-6180-2011-57},
  number = {1},
  journal = {EURASIP Journal on Advances in Signal Processing},
  publisher = {Springer Science and Business Media LLC},
  author = {Shrestha,  Raju and Mansouri,  Alamin and Hardeberg,  Jon Yngve},
  year = {2011},
  month = sep 
}

@inproceedings{Park2007,
  title = {Multispectral Imaging Using Multiplexed Illumination},
  url = {http://dx.doi.org/10.1109/ICCV.2007.4409090},
  DOI = {10.1109/iccv.2007.4409090},
  booktitle = {2007 IEEE 11th International Conference on Computer Vision},
  publisher = {IEEE},
  author = {Park,  Jong-Il and Lee,  Moon-Hyun and Grossberg,  Michael D. and Nayar,  Shree K.},
  year = {2007},
  pages = {1–8}
}

@article{Ohsawa2004,
  title = {Six Band HDTV Camera System for Spectrum-Based Color Reproduction},
  volume = {48},
  ISSN = {1943-3522},
  url = {http://dx.doi.org/10.2352/J.ImagingSci.Technol.2004.48.2.art00003},
  DOI = {10.2352/j.imagingsci.technol.2004.48.2.art00003},
  number = {2},
  journal = {Journal of Imaging Science and Technology},
  publisher = {Society for Imaging Science & Technology},
  author = {Ohsawa,  Kenro and Ajito,  Takeyuki and Komiya,  Yasuhiro and Fukuda,  Hiroyuki and Haneishi,  Hideaki and Yamaguchi,  Masahiro and Ohyama,  Nagaaki},
  year = {2004},
  month = mar,
  pages = {85–92}
}

@article{Imai2000,
  title = {A Comparative Analysis of Spectral Reflectance Estimated in Various Spaces Using a Trichromatic Camera System},
  volume = {44},
  ISSN = {1943-3522},
  url = {http://dx.doi.org/10.2352/J.ImagingSci.Technol.2000.44.4.art00003},
  DOI = {10.2352/j.imagingsci.technol.2000.44.4.art00003},
  number = {4},
  journal = {Journal of Imaging Science and Technology},
  publisher = {Society for Imaging Science & Technology},
  author = {Imai,  Francisco H. and Berns,  Roy S. and Tzeng,  Di-Y.},
  year = {2000},
  month = jul,
  pages = {280–287}
}

@article{Huang2022,
  title = {Spectral imaging with deep learning},
  volume = {11},
  ISSN = {2047-7538},
  url = {http://dx.doi.org/10.1038/s41377-022-00743-6},
  DOI = {10.1038/s41377-022-00743-6},
  number = {1},
  journal = {Light: Science \& Applications},
  publisher = {Springer Science and Business Media LLC},
  author = {Huang,  Longqian and Luo,  Ruichen and Liu,  Xu and Hao,  Xiang},
  year = {2022},
  month = mar 
}

@article{Zhang2025,
  title = {A review of the application of UAV multispectral remote sensing technology in precision agriculture},
  volume = {12},
  ISSN = {2772-3755},
  url = {http://dx.doi.org/10.1016/j.atech.2025.101406},
  DOI = {10.1016/j.atech.2025.101406},
  journal = {Smart Agricultural Technology},
  publisher = {Elsevier BV},
  author = {Zhang,  Shuang and Wang,  Xiaorui and Lin,  Hong and Dong,  Yueyu and Qiang,  Zhenping},
  year = {2025},
  month = dec,
  pages = {101406}
}

@article{Su2018,
  title = {Multispectral Imaging for Plant Food Quality Analysis and Visualization},
  volume = {17},
  ISSN = {1541-4337},
  url = {http://dx.doi.org/10.1111/1541-4337.12317},
  DOI = {10.1111/1541-4337.12317},
  number = {1},
  journal = {Comprehensive Reviews in Food Science and Food Safety},
  publisher = {Wiley},
  author = {Su,  Wen‐Hao and Sun,  Da‐Wen},
  year = {2018},
  month = jan,
  pages = {220–239}
}

@article{Ilianu2023,
  title = {Multispectral Imaging for Skin Diseases Assessment—State of the Art and Perspectives},
  volume = {23},
  ISSN = {1424-8220},
  url = {http://dx.doi.org/10.3390/s23083888},
  DOI = {10.3390/s23083888},
  number = {8},
  journal = {Sensors},
  publisher = {MDPI AG},
  author = {Ilișanu,  Mihaela-Andreea and Moldoveanu,  Florica and Moldoveanu,  Alin},
  year = {2023},
  month = apr,
  pages = {3888}
}

@article{Jones2020,
  title = {Understanding multispectral imaging of cultural heritage: Determining best practice in MSI analysis of historical artefacts},
  volume = {45},
  ISSN = {1296-2074},
  url = {http://dx.doi.org/10.1016/j.culher.2020.03.004},
  DOI = {10.1016/j.culher.2020.03.004},
  journal = {Journal of Cultural Heritage},
  publisher = {Elsevier BV},
  author = {Jones,  Cerys and Duffy,  Christina and Gibson,  Adam and Terras,  Melissa},
  year = {2020},
  month = sep,
  pages = {339–350}
}

@article{Qin2013,
  title = {Hyperspectral and multispectral imaging for evaluating food safety and quality},
  volume = {118},
  ISSN = {0260-8774},
  url = {http://dx.doi.org/10.1016/j.jfoodeng.2013.04.001},
  DOI = {10.1016/j.jfoodeng.2013.04.001},
  number = {2},
  journal = {Journal of Food Engineering},
  publisher = {Elsevier BV},
  author = {Qin,  Jianwei and Chao,  Kuanglin and Kim,  Moon S. and Lu,  Renfu and Burks,  Thomas F.},
  year = {2013},
  month = sep,
  pages = {157–171}
}

\end{document}